\documentclass[prl,twocolumn,amsmath,amssymb,showpacs,superscriptaddress]{revtex4-2}
\pdfoutput=1
\UseRawInputEncoding

\usepackage{graphicx}
\usepackage{epstopdf, epsfig}
\usepackage[utf8]{inputenc}
\usepackage[T1]{fontenc}
\usepackage{amsmath}
\usepackage[usenames,dvipsnames]{color}
\usepackage{dcolumn}% Align table columns on decimal point
\usepackage[FIGTOPCAP]{subfigure}
\usepackage[all]{xy}

\def \bea{\begin{eqnarray}}
	\def \eea{\end{eqnarray}}

\definecolor{mygray}{gray}{0.8}
\definecolor{plum}{rgb}{.5,0,1}
\usepackage{float}
\usepackage{mathtools}
\usepackage{amsmath,amssymb,amsbsy}
\usepackage{bbm,bm,mathrsfs,yfonts}
\usepackage[capitalize]{cleveref}
\usepackage{xcolor}
\usepackage[export]{adjustbox}
\usepackage[normalem]{ulem} %to strike the words

\newcommand\redsout{\bgroup\markoverwith{\textcolor{red}{\rule[0.5ex]{2pt}{0.4pt}}}\ULon}
\newcommand{\be}{\begin{equation}}
\newcommand{\ee}{\end{equation}}

\usepackage{xcolor}
\usepackage{soul}

\begin{document}

\setstcolor{red}

\title[Direct Microstability Optimization of Stellarator Devices]{Direct Microstability Optimization of Stellarator Devices}

\author{R. Jorge}
\address{Instituto de Plasmas e Fusão Nuclear, Instituto Superior Técnico, Universidade de Lisboa, 1049-001 Lisboa, Portugal}
\address{Department of Physics, University of Wisconsin-Madison, Madison, Wisconsin 53706, USA}
\email{rogerio.jorge@wisc.edu}

\author{W. Dorland}
\address{Department of Physics, University of Maryland, College Park, MD 20742, USA}
\address{Institute for Research in Electronics and Applied Physics, University of Maryland, College Park, MD 20742, USA}
\address{Princeton Plasma Physics Laboratory, Princeton, NJ 08543, USA}

\author{P. Kim}
\address{Department of Physics, University of Maryland, College Park, MD 20742, USA}

\author{M. Landreman}
\address{Institute for Research in Electronics and Applied Physics,
University of Maryland, College Park, MD 20742, USA}

\author{N. R. Mandell}
\address{Princeton Plasma Physics Laboratory, Princeton, NJ 08543, USA}

\author{G. Merlo}
\address{Oden Institute for Computational Engineering and Sciences, The University of Texas at Austin, Austin, TX 78712, United States of America}

\author{T. Qian}
%\address{%Department of Astrophysical Sciences, 
%Princeton University, Princeton, NJ 08543, USA}
\address{Princeton Plasma Physics Laboratory, Princeton, NJ 08543, USA}

\begin{abstract}
Turbulent transport is regarded as one of the key issues in magnetic confinement nuclear fusion, both for tokamaks in stellarators.
In this work, we show that a significant decrease in {a microstability-based proxy, as opposed to a geometric one, for the} turbulent heat flux{, namely the quasilinear heat flux,} can be obtained in an efficient manner by coupling stellarator optimization with linear gyrokinetic simulations.
This is accomplished by computing the quasi-linear heat flux at each step of the optimization process, as well as the deviation from quasisymmetry, and minimizing their sum, leading to a balance between neoclassical and {the} turbulent transport {proxy}.
\end{abstract}

\maketitle

\vspace{.4cm}{\section{Introduction}}

The stellarator is a class of fusion devices that can run in a steady state due to the flexible shaping of the confining magnetic field and the absence of current-driven instabilities \cite{Helander2014}.
This inherent flexibility allows stellarators to be optimized for a wide range of parameters, with W7-X \cite{Geiger2015} and HSX \cite{Anderson1995} examples of the experimental realizations of such optimization studies.
One of the main difficulties in stellarator optimization is the balance between neoclassical transport, driven by trapped particles in a low collisionality regime, and turbulent transport in the core, the latter being driven largely by microinstabilities \cite{Jenko2005}.
As an example, while W7-X has successfully shown decreased neoclassical transport, as in the tokamak case, turbulent transport is still limiting its energy confinement time \cite{Beidler2021}.
Indeed, turbulent transport is currently one of the main limiting factors in the performance of magnetic confinement fusion devices.
Although we are currently able to predict turbulent transport in the core with {moderate} accuracy, we still need an effective way to design machines with low turbulent transport, specifically, microinstability-driven turbulent transport \cite{Jenko2005,Holland2011}.
%
% A notable example of turbulence reduction via changes in the magnetic field equilibrium is the doubling of the confinement time in the TCV tokamak by reversing the triangularity of the poloidal cross-section of the flux surfaces \cite{Camenen2007a}.

In this work, we directly target both (1) the reduction of the microinstability drive of turbulent transport, and (2) the trapped particle losses driving neoclassical transport.
As we show here, depending on the weights used {in the optimization}, a balance between the two types of transport is possible, and magnetic field equilibria with reduced quasilinear heat flux can be achieved.
While previous optimization methods targeted cost functions based on proxy functions relying solely on the properties of the magnetic field equilibrium \cite{Mynick2010, Xanthopoulos2014, Roberg-Clark2022}, here, we do not rely on {geometry-based} proxy functions to evaluate microstability and directly solve the gyrokinetic equation at each iteration.
To our knowledge, this is the first study where gyrokinetic calculations are performed within the optimization at every iteration.
We note that, while in this study a particular numerical tool is used to solve the gyrokinetic equation and extract its linear growth rates and eigenfunctions, the direct optimization method employed here can be applied to other equilibrium and gyrokinetic codes which model different instability types and can easily be generalized to directly optimize turbulent heat and particle fluxes.

% An example of the reduction of the ITG instability drive leading to lower turbulent transport was reported in Ref. \cite{Xanthopoulos2014} with the design of the MPX stellarator.
% %
% Here, the cost function used was the negative part of the radial covariant component of the curvature, $\kappa_r^{-}=\left(\mathbf B \times \nabla B \cdot \mathbf k_\perp\right)^{-}/B^2$.
% %
% This reduces the ITG growth rate by either reducing the unfavorable ("bad") curvature or by increasing the distance between adjacent flux surfaces in locations where the curvature is strong.
% %
% Another example of a proxy function to reduce turbulence is the one used in Ref. \cite{Mynick2010} where a quasi-linear estimate for the ion conductivity was used to reduce the associated transport.
%
% Finally, we mention here the recent approach by Ref. \cite{Roberg-Clark2022} where the linear onset of ITG modes ("critical gradient") was increased using as a proxy function the drift curvature $\omega_d(l)$.

The framework used in this work to assess microstability is gyrokinetics.
This is considered to be one of the main tools to assess the stability of fusion devices at spatial scales on the order of, or smaller than, the ion gyroradius $\rho_i$ and at frequencies lower than the ion gyrofrequency $\Omega_i$ \cite{Catto1978a,Antonsen1980,Frieman1980} and is usually regarded as the most complete, yet numerically efficient, framework to treat strongly magnetized plasmas \cite{Helander2021a}.
For this reason, we quantify instabilities in this work by solving the linearized gyrokinetic equation to obtain the growth rates and corresponding eigenfunctions associated with the underlying instabilities.
The instability studied here is the ion-temperature-gradient (ITG) mode, which is commonly regarded as one of the most transport-relevant electrostatic instabilities in tokamaks and stellarators \cite{Horton1999,Helander2013}.

The ITG mode develops on the ion gyroscale and is widely recognized as one of the main candidates to explain the experimental observations of anomalous ion heat turbulent transport in the core of fusion devices \cite{Garbet2004} as corroborated by numerical simulations \cite{Hahm1998,Romanelli1998,Plunk2014}.
The study of solely ion-driven instabilities such as the ITG mode is performed by removing the fast electron dynamics induced by electron inertia, meaning that the electron density in our model follows the perturbed electrostatic potential $\phi$ via the Boltzmann response.
The ITG mode can also deteriorate plasma confinement in electron heating scenarios by creating ion temperature clamping {in low-power W7-X ECRH-only discharges without the use of the pellet injection or the boron dropper}, preventing the heating of ions in the plasma core above 2 keV \cite{Beurskens2021}.
While ITG can drive a certain amount of electron heat flux (usually proportional to the driven ion heat flux), contributions to the electron heat flux from trapped electron modes are expected if the ITG is not sufficient to drive the electron heat flux imposed by the applied heating power \cite{Ryter2019}.
%
% This is due to the additional free energy present in the system fueled by the density and temperature gradients of the electrons that can lead to instabilities both at the $k_y \sim 1$ scale (TEM) and $k_y \gg 1$ scale (ETG).
% %
% A transition from ITG to TEM, i.e., a mode propagation change from the ion $\Re(\omega)>0$ to the electron $\Re(\omega)<0$ diamagnetic direction, is expected at $k_y \sim 0.1-1$.
% %
% Furthermore, while the ITG mode is stabilized by finite Larmor radius (FLR) effects, $k_\perp>1$, the TEM growth rate is expected to increase with $k_\perp$.
%
% In this work, electron density and temperature gradients as it greatly increases the numerical cost of the optimization, the optimization method shown here can easily be generalized to include electron-driven instabilities by adding electron inertia to the equations being solved by the gyrokinetic simulations.

As an optimization criterion to stabilize the underlying unstable modes, we choose our objective function $J$ to be a quasi-linear estimate \cite{Jenko2005,Fable2009} for the heat flux $f_\text{Q}$ with the growth rates $\gamma$ and associated eigenfunctions calculated using linear gyrokinetic simulations.
The method employed here can also be used with nonlinear gyrokinetic or fluid codes by replacing the quasi-linear estimate $f_{\text{Q}}$ with the nonlinear heat flux $Q$ computed by the respective code.
This differs from previous objective functions that also targeted the reduction of turbulent transport such as Refs. \cite{Mynick2010,Proll2015b,Roberg-Clark2022} as these are solely based on properties of the equilibrium magnetic field {and not on direct evaluations of gyrokinetic simulations}, therefore requiring more assumptions about the geometry and the underlying modes.
We note that, while the peak growth rate $\gamma$ could be chosen as the objective function, we found it not to be a reliable indicator of the nonlinear heat flux due to the importance of smaller wave numbers \cite{Pueschel2016}.
The reduction of neoclassical transport is performed by targeting quasisymmetry, an invariance of the magnetic field strength $B$ that guarantees confinement of the collisionless trajectories up to a threshold energy \cite{Nuhrenberg1988a}.
As shown in Ref. \cite{Landreman2022}, it is indeed possible to design stellarators with precise quasisymmetry and achieve unprecedented levels of collisionless particle confinement and collisional transport for a thermal plasma.
We leverage such findings by adding to the objective function $J$ the term $f_{\text{QS}}$ defined in Eq. (1) of Ref. \cite{Landreman2022} as it is already in a form ready to be used in a least squares optimization method and, unlike previous optimization metrics, it does not require the calculation of Boozer coordinates \cite{Boozer1981} at each optimization step.
{We note that the quasisymmetry metric $f_{\text{QS}}$ is, in fact, a measure of the deviation of quasisymmetry. Therefore, during optimization, we seek to minimize $f_{\text{QS}}$.}

\vspace{.4cm}{\section{Method}}

The growth rates $\gamma$ are calculated using the GS2 code \cite{Kotschenreuther1995,Dorland2000a,Baumgaertel2011} that solves the gyrokinetic equation
\begin{align}
    \frac{d h}{d t}+v_\parallel \mathbf b \cdot \nabla h + \mathbf v_d \cdot \nabla h = C + \frac{q F_0}{T_0}\frac{\partial \chi}{\partial t},
\label{eq:gk}
\end{align}
coupled with the quasineutrality condition and Ampère's law.
In \cref{eq:gk}, $F_0$ is the equilibrium distribution function, $h=f-F_0(1-q \phi/T_0)$ is the non-adiabatic part of the distribution function $f$, $\chi=\left<\phi - \mathbf v \cdot \mathbf A/c\right>$ is the gyroaveraged gyrokinetic potential with $\phi$ the electrostatic potential and $\mathbf A$ the magnetic vector potential, $d_t=\partial_t + (c/B)[\chi, \cdot]$ is the total time derivative with $[\chi,h]=\partial \chi/\partial \mathbf R \times \partial h/\partial \mathbf R \cdot \mathbf b$ the Poisson bracket, $\mathbf v_d = (\mathbf b / 2\Omega_0)\times(2v_\parallel^2 \mathbf b \cdot \nabla \mathbf b + v_\perp^2 \nabla B_0/B_0)$ is the drift velocity with $B_0$ the equilibrium magnetic field, $\Omega_0=q B_0/m$ is the gyrofrequency with $q$ the particle's charge and $m$ its mass, and $C$ is the gyroaveraged collision operator.
To focus on ITG modes, we reduce the gyrokinetic equation to its linear electrostatic flux-tube limit employing the adiabatic electron approximation, solve it as an initial value problem, and fit the temporal evolution of the absolute value of the electrostatic potential to an exponential of the form $\exp(\gamma t)$ with $\gamma$ the growth rate.

The GS2 {code employs} a Fourier decomposition of the physical quantities in the $x$ (normal) and $y$ (binormal) directions with $\mathbf k_\perp = k_x \nabla x + k_y \nabla y$ the perpendicular wave-vector and $\mathbf r = (x,y,z)$ the spatial coordinates.
Such coordinates are based on the Clebsch representation for the equilibrium magnetic field $\mathbf B = \nabla \psi \times \nabla \alpha$ with $\psi$ the toroidal magnetic flux and $\alpha$ a field line label on the flux surface {defined in Boozer coordinates $(\psi,\theta,\varphi)$ as $\alpha=\theta-\iota \varphi$} with $\iota=d\theta/d\varphi$ \cite{Helander2014}.
The coordinates are then defined as $x=a \sqrt{s}$ with $a$ a reference length taken here as the effective minor radius of the stellarator (see definition in page 12 of Ref. \cite{Landreman2019b}) and $s=\psi/\psi_b$ where $\psi_b$ is the value of $\psi$ at the outermost flux surface (otherwise known as plasma boundary), $y=x \alpha$ and $z$ a dimensionless coordinate along the field line.

The density and temperature profiles are assumed to decay exponentially in $x$ with a constant scale length $L_{n}/a= - d \ln n/d x$ and $L_{T}/a= - d \ln T/d x$, respectively.
For this work, we simulate flux tubes with $s=0.25$ and linearly evolve modes with $k_x=\alpha=0$ with a grid of ten values of $0.3\le k_y\le 3.0$.
However, we note that while the choice $\alpha=0$ is expected to lead to a peak of the modes mainly on the outboard side \cite{Plunk2014} where the choice $k_x=0$ is likely to yield growth rates at or close to the maximum one, this hypothesis{, when applied to the design of new devices, should be} corroborated \textit{a posteriori} via nonlinear simulations at finite $k_y$ and $k_x$.
The density and temperature profiles used are $a/L_n=1$ and $a/L_T=3$, respectively, with $a$ the minor radius, leading to $\eta = L_T/L_n=3$, a typically used value in cyclone base case scenarios \cite{Dimits2000}.
The remaining input parameters for GS2 were obtained by performing convergence tests for the initial and final equilibria of each optimization in a similar fashion to Ref. \cite{Jorge2021}.
{In particular, we use a total of 5 poloidal turns, 151 points along the field line and 35 points along the velocity pitch angle variable $\lambda$.}

We then write the quasi-linear estimate $f_\text{Q}$ for the heat flux using a mixing length saturation rule \cite{Mariani2018}
\begin{equation}
    f_{\text{Q}}=\sum_{k_y}\frac{\gamma(k_y)}{\left<k_\perp^2{(k_y)}\right>},
\label{eq:quasilineareq}
\end{equation}
where the sum is taken over all values of $k_y$ used in the simulation, and $\left<k_\perp^2\right>=\int g^{yy}k_y^2 |\hat{\phi}|^2 \sqrt{g} dz/\int |\hat{\phi}|^2 \sqrt{g} dz$ is the flux-surface average of the squared perpendicular wave number with $|\hat{\phi}|$ the amplitude of the mode and $\sqrt{g}$ the Jacobian.
As shown in \cite{Jenko2005}, such quasi-linear transport model overcomes some of the limitations of the heuristic mixing length estimate $\gamma/k_y^2$ by taking into account the extension of the toroidal modes along $z$.

The components of the magnetic field equilibrium relevant to compute both the geometric coefficients entering \cref{eq:gk} and the quasisymmetry objective function $f_{\text{QS}}$ are calculated using the Variational Moments Equilibrium Code (VMEC) \cite{Hirshman1983}.
VMEC finds a solution of ideal magnetohydrodynamics (MHD) equation $\mathbf J \times \mathbf B = \nabla P$, with $\mathbf J = \nabla \times \mathbf B/\mu_0$ the plasma current and $P$ the plasma pressure, by applying a variational method to the integral form of the MHD equations.
We run VMEC in fixed-boundary mode where an MHD equilibrium is obtained by specifying the last closed flux surface $S$ as boundary condition parametrized by two variables $(\theta,\phi)$, namely an angle-like variable termed the poloidal angle $\theta$ and the standard cylindrical toroidal angle $\phi$, respectively, allowing us to define $S=[R(\theta,\phi)\cos(\phi), R(\theta,\phi)\sin(\phi), Z(\theta,\phi)]$ where $R=\sum_{m=0}^{M_{\mathrm{pol}}}\sum_{n=-N_{\mathrm{tor}}}^{N_{\mathrm{tor}}}\mathrm{RBC}_{m,n}\cos(m\theta-n_{\text{fp}}n\phi)$ and $Z=\sum_{m=0}^{M_{\mathrm{pol}}}\sum_{n=-N_{\mathrm{tor}}}^{N_{\mathrm{tor}}}\mathrm{ZBS}_{m,n}\sin(m\theta-n_{\text{fp}}n\phi)$.
In the Fourier decompositions of $R$ and $Z$, the $\sin$ and $\cos$ terms are set to zero, respectively, to ensure stellarator-symmetry \cite{Dewar1998} and $n_{\text{fp}}$ is the toroidal periodicity of the equilibrium.
While stellarator symmetry is, in principle, not necessary, it allows us to restrict the total number of degrees of freedom while still providing adequate optimized solutions.
Given a plasma boundary $S$, a plasma pressure profile $P(\psi)$, and a net toroidal current $I$, VMEC seeks a solution of the ideal MHD system of equations with nested closed flux surfaces.

The optimization algorithm, the calculation of the quasisymmetry objective function, and the calculation of the geometric coefficients that enter the gyrokinetic equation are performed using the SIMSOPT code \cite{Landreman2021b}.
The independent variables for optimization are the boundary Fourier coefficients $\{\mathrm{RBC}_{m,n},\mathrm{ZBS}_{m,n}\}$ excluding $\mathrm{RBC}_{0,0}$ which is set to one to fix the spatial scale.
We use as initial condition the precise quasi-helically (QH) symmetric configuration of Ref. \cite{Landreman2022} with $n_{fp}=4$ and set to zero all surface modes with {$m\ge1$ and $|n|\ge1$}.
Finally, in order to keep the same aspect ratio as the provided initial condition to the optimization, we add to the objective function $J$ the term $f_A=(A-A^{*})^2$ where $A$ is the aspect ratio of the configuration computed at each iteration and $A^{*}=8$ the target (initial) aspect ratio.
The final objective function is then given by
\begin{equation}
    J=\omega_{f_{\text{Q}}} f_{\text{Q}}+f_{\text{QS}}+f_A,
\end{equation}
where $\omega_{f_{\text{Q}}}$ is the weight given to the minimization of $f_{\text{Q}}$.

\begin{figure}
    \centering
    \includegraphics[width=.4\textwidth, trim={0.45cm 0.45cm 0.5cm 0.1cm}, clip]{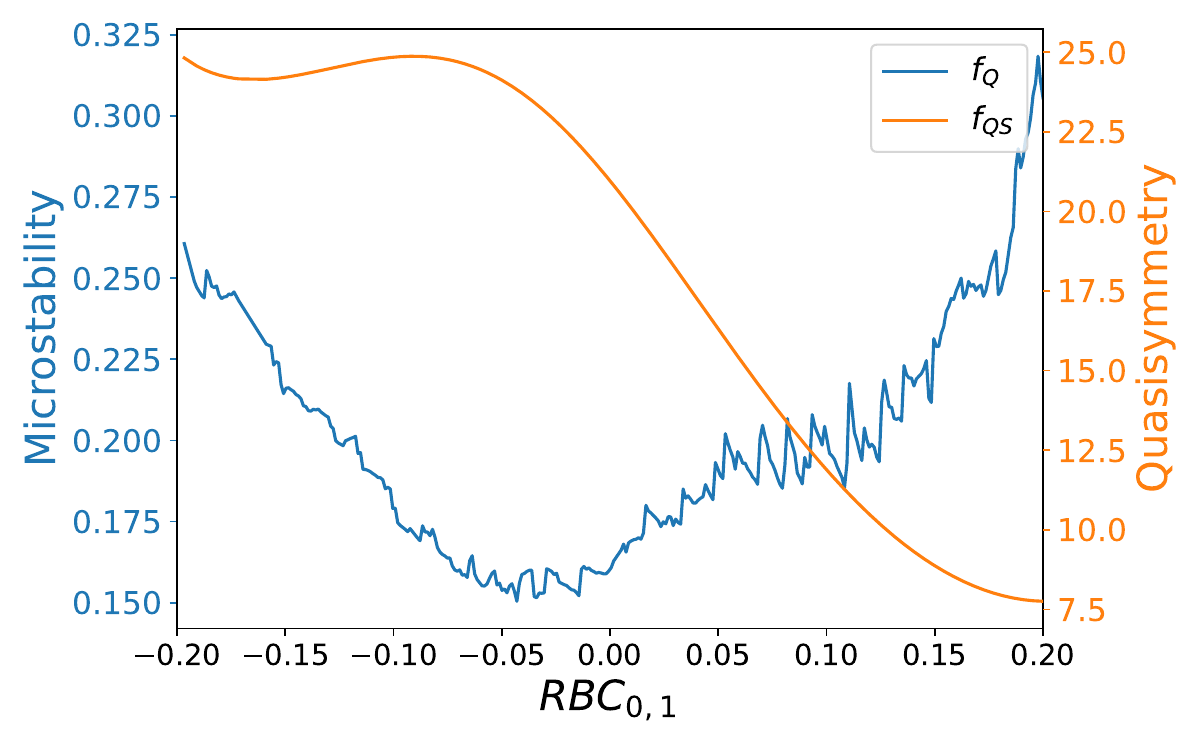}
    \caption{Quasi-linear estimate $f_\text{Q}$ (noisy, lower line) and quasisymmetry cost function $f_{QS}$ (smooth, uppwer line) as a function of the surface shape parameter $\mathrm{RBC}_{0,1}$ for the quasi-helically symmetric configuration used as initial conditions for the optimization.}
    \label{fig:ITG_scan}
\end{figure}

\vspace{.4cm}{\section{Results}}

We first perform a parameter space scan to guide the choice of our optimization algorithm.
Namely, we assess the dependence of the quasisymmetry cost function $f_{\text{QS}}$ and the quasi-linear estimate $f_{\text{Q}}$ on the geometry of the magnetic field by performing a scan on the $\mathrm{RBC}_{0,1}$ boundary mode in the range $(-0.2,0.2)${, a range of values where VMEC {is} able to yield converged } {with a tolerance of $\mathrm{FTOL}=10^{-14}$ with 131 radial points within 7000 iterations} on the initial QH configuration where $\mathrm{RBC}_{0,1}=0.180$.
This is shown in \cref{fig:ITG_scan} with the value of $f_{\text{Q}}$ illustrating microstability in blue and the value of $f_{\text{QS}}$ illustrating quasisymmetry in orange.
We find that while there are only two local minima of $f_{\text{QS}}$, the cost function $f_{\text{Q}}$ contains many local minima with its global minimum at a different location than the global minimum of $f_{\text{QS}}$.
Therefore, when performing optimization studies on both quasisymmetry and microstability, a compromise on the level of quasisymmetry is expected.
Given the fact that the local minima of $f_{\text{Q}}$ and $f_{\text{QS}}$ appear to be well separated, and that the roughness of the cost function $f_{\text{Q}}$ does not lead to large jumps in $f_{\text{Q}}$ for small changes of $\mathrm{RBC}_{0,1}$, we use a local optimization algorithm instead of a global optimization algorithm.
This leads to a more efficient optimization method but it may not result in the global minimum of the objective function found.
However, as it is shown below, the solution found is able to fulfill our desired criteria.
The optimization is carried out using the Python library Scipy 
\cite{Virtanen2020}
for nonlinear least-squares minimization by employing the Levenberg-Marquardt method.
Gradients are computed using finite differences, with MPI for concurrent function evaluations where, at each iteration, both the magnetic field inside the boundary (using VMEC) and the gyrokinetic simulations (using GS2) are computed within the SIMSOPT framework.
In order to steer the optimizer away from the many local minima present in $f_{\text{Q}}$ and still be able to use finite differences, we use a large value for the relative $\Delta_r$ and absolute $\Delta_a$ step sizes and perform {two} extra optimization to refine the minimum found.
{
Due to the smallness of the degrees of freedom with increasing Fourier modes, for a given maximum mode $M_{\mathrm{pol}}$, we use $\Delta_r=0.015/M_{\mathrm{pol}}$ and $\Delta_a=$min$[0.003,(M_{\mathrm{pol}}/5)\times10^{-M_{\mathrm{pol}}}]$ in the first optimization, then decreasing $\Delta_r$ and $\Delta_a$ by a factor of 10 and 300 for the two ensuing optimizations, respectively.
}
We note that the study in \cref{fig:ITG_scan} is performed by varying a single parameter and the extrapolation of such findings to other optimization parameters would need a multi-dimensional survey.
Therefore, an \textit{a posteriori} examination of the growth rates and is performed to assess the effectiveness of this optimization procedure.

\begin{figure}
    \includegraphics[width=.23\textwidth, trim={0.20cm 1.4cm 0.1cm 0.5cm}, clip]{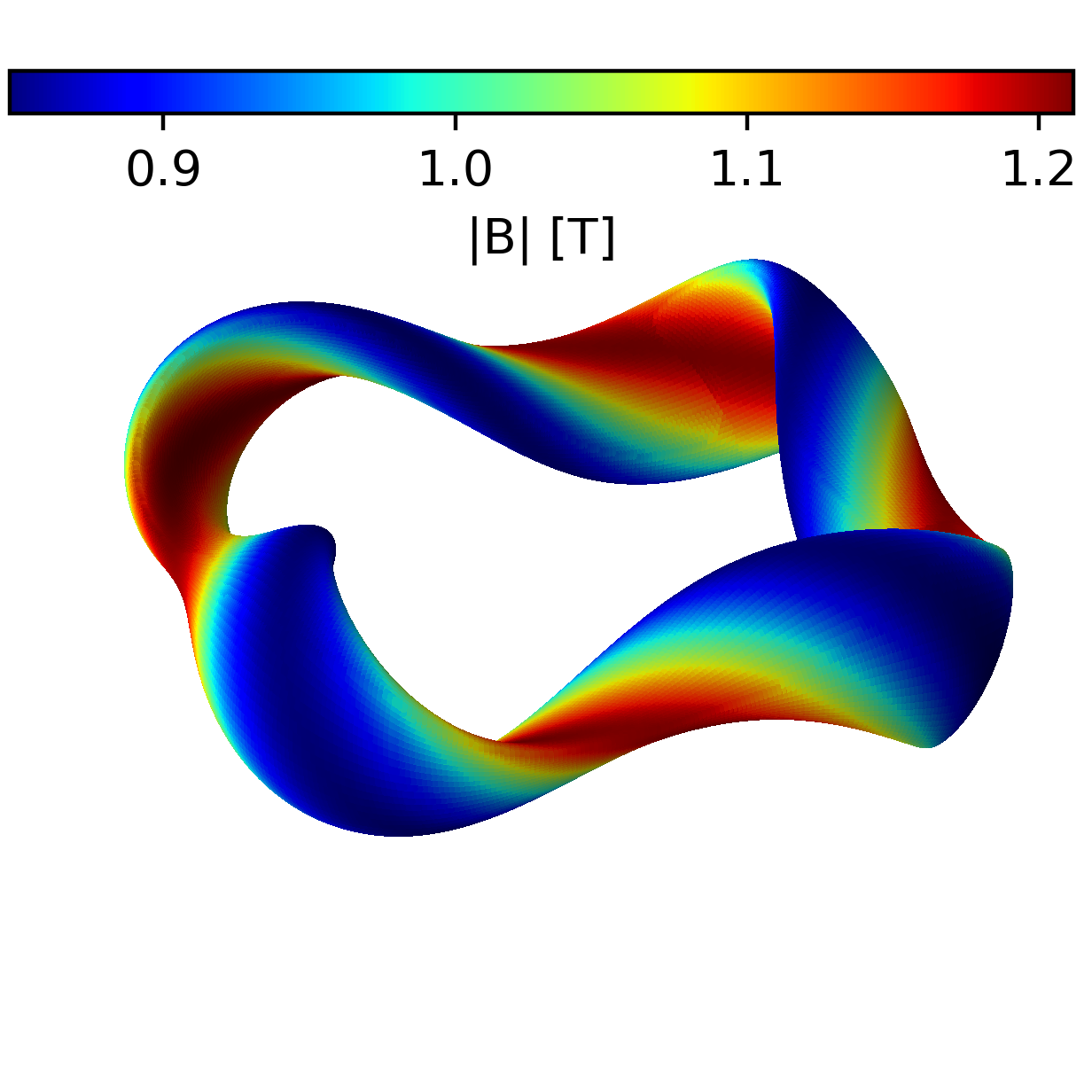}
    \includegraphics[width=.23\textwidth, trim={0.20cm 1.4cm 0.1cm 0.5cm}, clip]{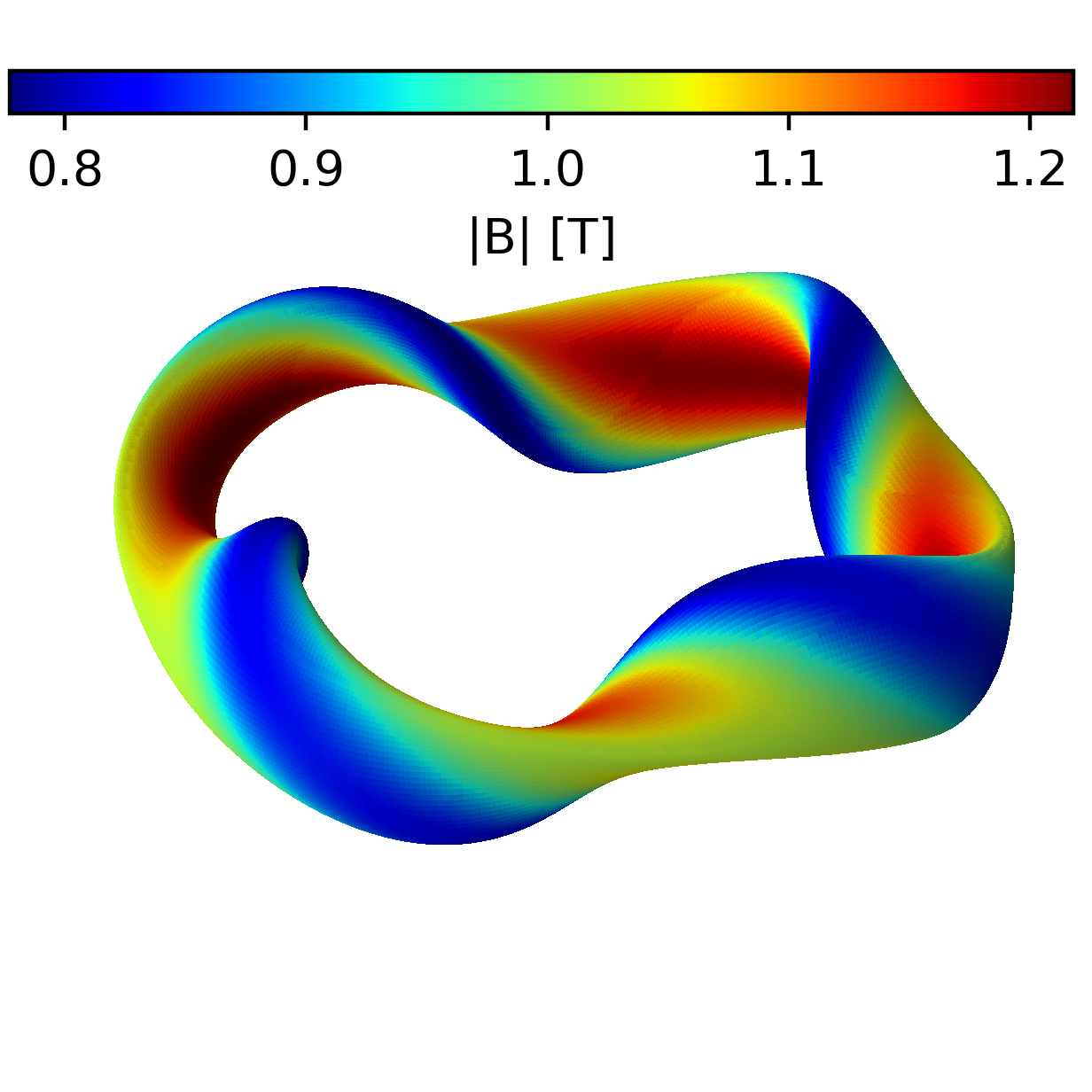}
    \includegraphics[width=.23\textwidth]{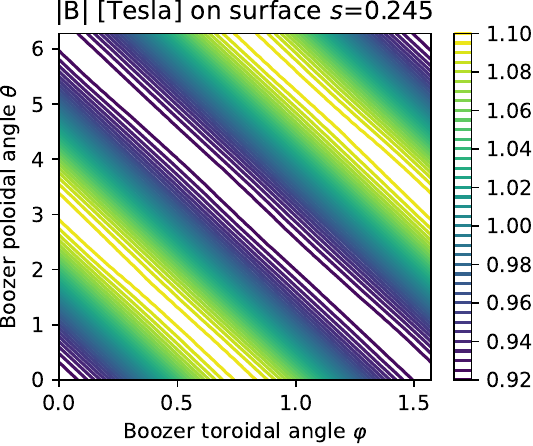}
    \includegraphics[width=.23\textwidth]{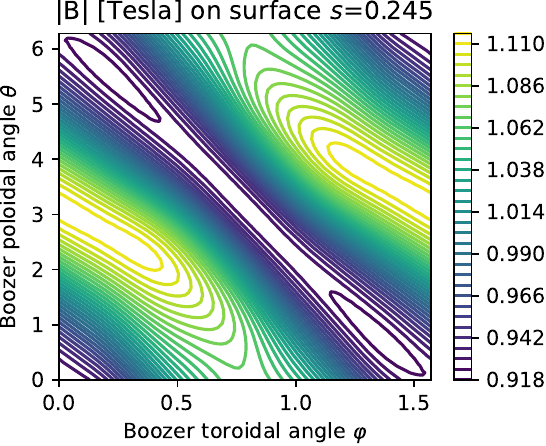}
    % \includegraphics[width=.23\textwidth,
    % trim={0.3cm 0.3cm 0.25cm 0.1cm}
    % , clip]{gs2_scan_weighted_gamma_onlyQS.pdf}
    % \includegraphics[width=.23\textwidth,
    % trim={0.3cm 0.3cm 0.25cm 0.1cm}
    % , clip]{gs2_scan_weighted_gamma.pdf}
    \caption{
    Top: Magnetic field configuration optimized for quasisymmetry with $\omega_{f_{\text{Q}}}=0$ (left) and the configuration optimized for quasisymmetry and microstability with $\omega_{f_{\text{Q}}}=10$ (right) where the color shows the magnetic field strength on that surface.
    {Bottom}: Field strength on an $s=0.245$ flux surface in Boozer coordinates for $\omega_{f_{\text{Q}}}=0$ (left) and $\omega_{f_{\text{Q}}}=10$ (right).
    }
    \label{fig:ITG_optimized_fig_QH_1}
\end{figure}

\begin{figure}
    % \includegraphics[width=.23\textwidth, trim={2.25cm 3.55cm 3.0cm 1.6cm}, clip]{wout_3Dplot_VMEC_onlyQS.png}
    % \includegraphics[width=.23\textwidth, trim={2.25cm 3.55cm 3.0cm 1.6cm}, clip]{wout_3Dplot_VMEC.png}
    % \includegraphics[width=.23\textwidth]{Boozxform_surfplot_2_onlyQS.pdf}
    % \includegraphics[width=.23\textwidth]{Boozxform_surfplot_2.pdf}
    % \includegraphics[width=.23\textwidth,
    % trim={0.3cm 0.3cm 0.25cm 0.1cm}
    % , clip]{gs2_scan_weighted_gamma_onlyQS.pdf}
    \includegraphics[width=.23\textwidth,
    trim={0.3cm 0.3cm 0.25cm 0.1cm}
    , clip]{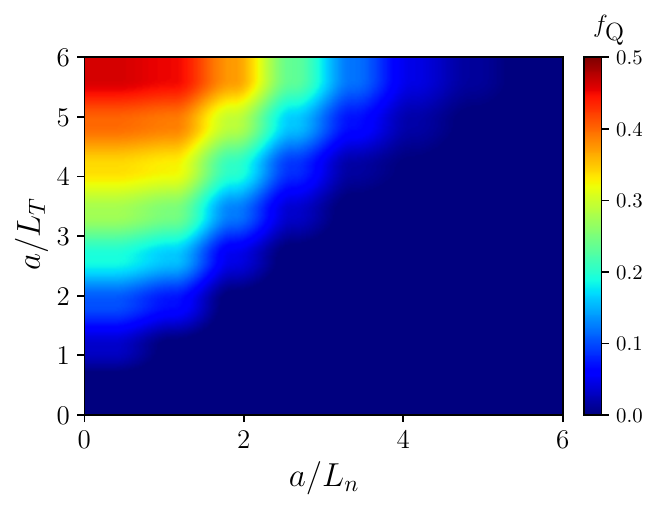}
    % \includegraphics[width=.23\textwidth,
    % trim={0.3cm 0.3cm 0.25cm 0.1cm}
    % , clip]{gs2_scan_weighted_gamma.pdf}
    \includegraphics[width=.23\textwidth,
    trim={0.3cm 0.3cm 0.25cm 0.1cm}
    , clip]{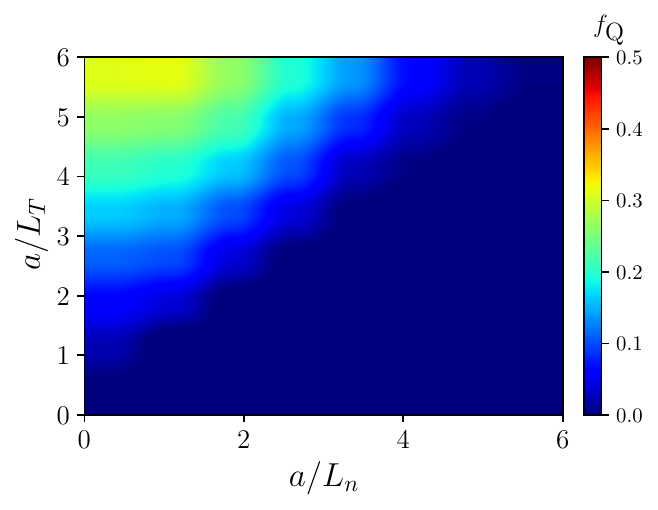}
    \includegraphics[width=.23\textwidth,
    trim={0.3cm 0.3cm 0.25cm 0.1cm}
    , clip]{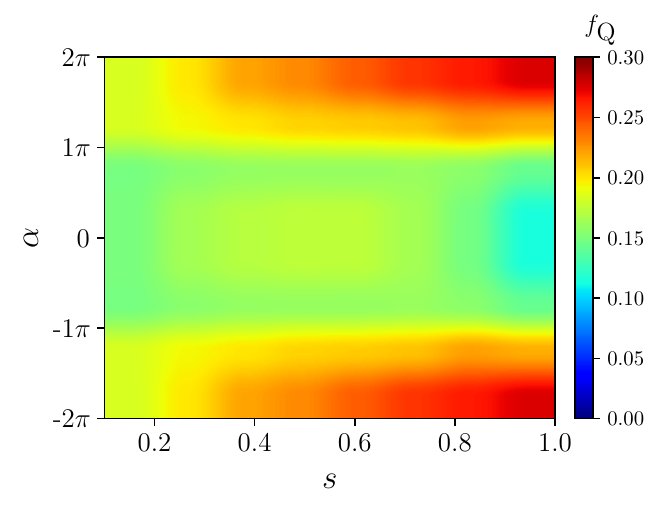}
    \includegraphics[width=.23\textwidth,
    trim={0.3cm 0.3cm 0.25cm 0.1cm}
    , clip]{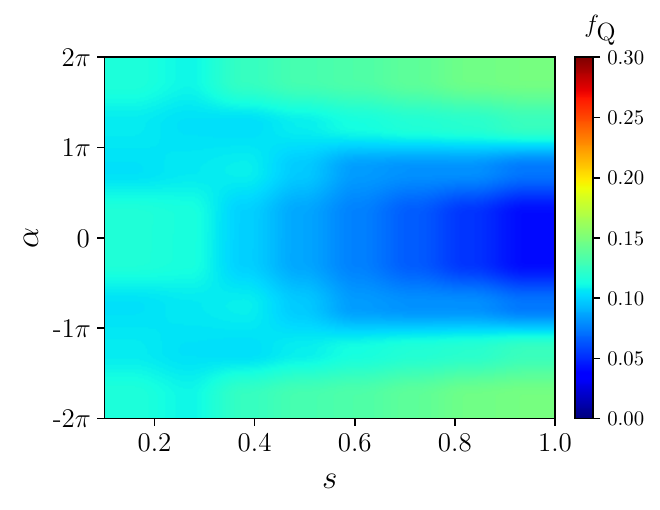}
    \caption{
    {Top: Comparison of the quasi-linear estimate for the heat flux between the configuration optimized for $\omega_{f_{\text{Q}}}=0$ (left) and $\omega_{f_{\text{Q}}}=10$ (right) for several profile scale lengths $L_n$ and $L_T$. {We note that the stability (dark blue color) observed in the lower right corners is a characteristic of the adiabatic electron assumption employed here.}
    Bottom: Similar comparison at fixed $a/L_{n}=1$ and $a/L_{T}=3$ for different field line labels $\alpha$ and surfaces $s$ corresponding to different fluxtubes.
    }
    }
    \label{fig:ITG_optimized_fig_QH_2}
\end{figure}

We perform {five} optimization studies.
The first targets only quasisymmetry by setting $\omega_{f_{\text{Q}}}=0$ and will be used as a benchmark case.
The resulting optimization is shown in \cref{fig:ITG_optimized_fig_QH_1} (left).
The {remaining studies} target both microstability and quasisymmetry simultaneously and set {$\omega_{f_{\text{Q}}}=0.1, 1, 10$ and 100}.
The resulting optimization {for the value of $\omega_{f_{\text{Q}}}=10$} is shown in \cref{fig:ITG_optimized_fig_QH_1} (right).
The overall optimization was done in an iterative fashion, namely, an optimization with large and smaller step sizes was performed at each of maximum surface Fourier modes $M_{\mathrm{pol}}=N_{\mathrm{tor}}=1, 2, 3, 4$.
A further increase of $M_{\mathrm{pol}}$ and $N_{\mathrm{tor}}$ resulted in negligible improvements in the objective function in the $\omega_{f_{\text{Q}}}=10$ case.
Therefore, the resulting configurations have only surface Fourier modes up to $m\le4$ and $|n|\le4$.
{The initial values of $f_{\text{QS}}$ and $f_{\text{Q}}$ are $1.412\times10^{-1}$ and $1.346\times10^{-1}$, respectively.}
{
We find that, for the cases of $\omega_{f_{\text{Q}}}=0, 0.1, 1, 10$ and 100, the quasisymmetry objective function increases from $f_{\text{QS}}=2\times10^{-4}$ to $4\times10^{-4}$, $1.11\times10^{-2}$, $1.171\times10^{-1}$ and $2.966\times10^{-1}$, respectively, while the microstability objective function decreases from $f_{\text{Q}}=1.92\times10^{-1}$ to $1.91\times10^{-1}$, $1.63\times10^{-1}$, $1.12\times10^{-1}$ and $1.01\times10^{-1}$, respectively.
}
In order to assess the deterioration in quasisymmetry, we show in \cref{fig:ITG_optimized_fig_QH_1} {(bottom)} the contours of the magnetic field strength at $s=0.245$ with the deviation from straight lines appearing in the second case (optimization for microstability).
On the other hand, the enhanced microstability properties of the resulting configuration {are} evident even at values of $L_n$ and $L_T$ outside the ones used for the optimization, {and for several flux tubes and radii,} see {\cref{fig:ITG_optimized_fig_QH_2}}.

\begin{figure}
    \includegraphics[width=.23\textwidth,
    trim={0.3cm 0.3cm 0.25cm 0.1cm}
    , clip]{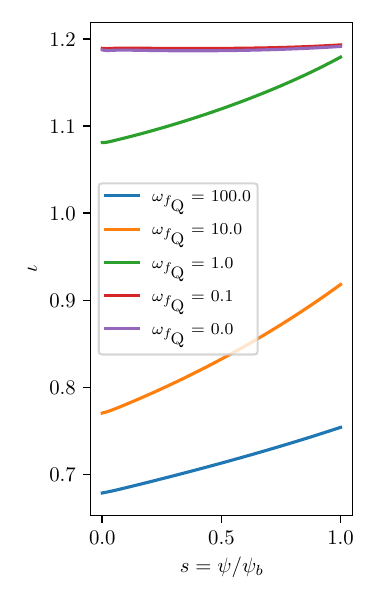}
    \includegraphics[width=.235\textwidth,
    trim={0.3cm 0.3cm 0.25cm 0.1cm}
    , clip]{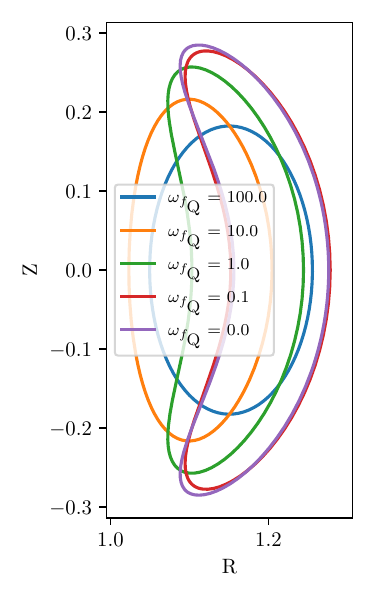}
    \includegraphics[width=.45\textwidth,
    trim={0.3cm 0.4cm 0.25cm 0.1cm}
    , clip]{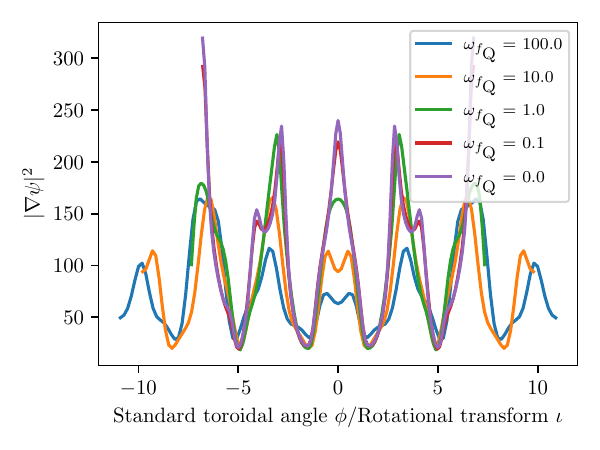}
    \caption{
    {Radial profiles of rotational transform $\iota$ (top left), cross-sectional cuts of the plasma boundary at cylindrical angle $\phi=0$ and metric tensor coefficient $g^{\psi\psi}=|\nabla \psi|^2$ along a field line (bottom) for the five optimization cases considered here at $\omega_{f_{\text{Q}}}=0, 0.1, 1, 10$ and 100. $R$ and $Z$ are shown in units of the major radius $R_0$, here taken as $R_0=1$m.
    }
    }
    \label{fig:ITG_optimized_properties}
\end{figure}

{The properties of the five different configurations obtained from the optimizations at $\omega_{f_{\text{Q}}}=0, 0.1, 1, 10$ and $100$ are now assessed.
We compare in \cref{fig:ITG_optimized_properties} the rotational transform $\iota$ (top left), cross-sections at a cylindrical angle $\phi=0$ (top right), and geometry coefficient $|\nabla \psi|^2$ (bottom) between the different configurations.
We find that, as microstability is favored with increasing $\omega_{f_{\text{Q}}}$, the rotational transform decreases and, comparing with $\omega_{f_{\text{Q}}}=0$ and $0.1$, there is a finite magnetic shear.
{We note that magnetic shear has been associated with turbulence reduction, see Refs. \cite{Xanthopoulos2016,Plunk2017,Faber2018}.}
Furthermore, it is found that the cross-sections become more circular { and, therefore, have less surface compression,} as $\omega_{f_{\text{Q}}}$ is increased.
Finally, as shown in \cref{fig:ITG_optimized_properties} (bottom), the maxima of the metric tensor coefficient $|\nabla \psi|^2$ are seen to decrease as the value of $\omega_{f_{\text{Q}}}$ increases.
This parameter {represents flux surface compression and} plays a role in ITG turbulence as it is the coefficient of the ion radial heat flux $Q_i\propto - |\nabla \psi|^2 d T_i/dx$ \cite{Mynick2010}.
}

\begin{figure}
    \includegraphics[width=.45\textwidth,
    trim={0.3cm 0.3cm 0.25cm 0.1cm}
    , clip]{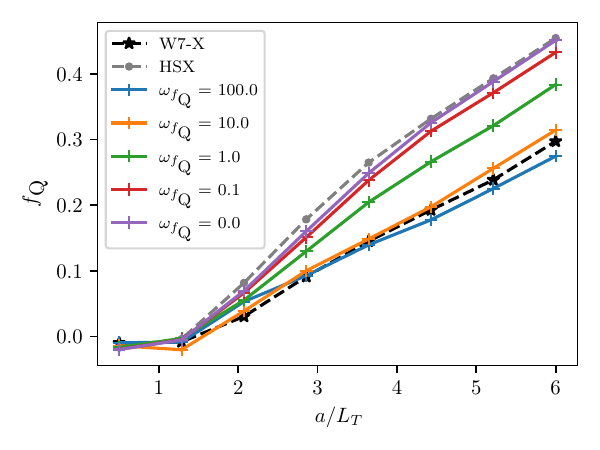}
    \caption{
    {Comparison of the quasi-linear estimate $f_Q$ from \cref{eq:quasilineareq} between the standard configuration of W7-X, HSX, and the five optimizations carried out here at $\omega_{f_{\text{Q}}}=0, 0.1, 1, 10$ and 100 as a function of the minor radius $a$ divided by the gradient scale length of the ion temperature $a/L_T$.
    }
    }
    \label{fig:ITG_optimized_properties_W7X}
\end{figure}

We now perform an \textit{a posteriori} assessment of the neoclassical {and quasi-linear transport} of the optimized configurations{,} {free-boundary} standard configuration of the W7-X device \cite{Beidler1990} {and the free-boundary configuration of the HSX device \cite{Anderson1995}. Both the W7-X and HSX files consist of the vacuum configurations taken from the public repository in Ref. \cite{vmec_equilibria}}.
{
We start with a comparison of the quasi-linear proxy $f_\text{Q}$ between the five different optimizations and the W7-X device.
This is shown in \cref{fig:ITG_optimized_properties_W7X}.
Here we find that a precise QH configuration with four field periods has a larger quasi-linear flux than W7-X.
A similar conclusion holds for the maximum growth rates and for different flux tubes.
However, the $f_\text{Q}$ decreases to levels similar to W7-X for $\omega_{f_{\text{Q}}}=10$, and reaches even lower values for $\omega_{f_{\text{Q}}}=100$.
Additionally, we note that the optimization performed here also led to {similar, or possibly higher,} $f_\text{Q}$ critical gradient values of $a/L_T$ for the case of $\omega_{f_{\text{Q}}}=10$ when compared with other values of $\omega_{f_{\text{Q}}}$.
}

\begin{figure}
    \centering
    % \includegraphics[width=.4\textwidth,
    % trim={0.50cm 0.45cm 0.35cm 0.35cm}
    % , clip]{neo_out_QH.pdf}
    % \includegraphics[width=.23\textwidth,
    % % trim={0.35cm 0.45cm 0.35cm 0.4cm}
    % % , clip
    % ]{GX_QH_heatFluxes.png}
    \includegraphics[width=.46\textwidth,
    trim={0.20cm 0.40cm 0.35cm 0.35cm}
    , clip]{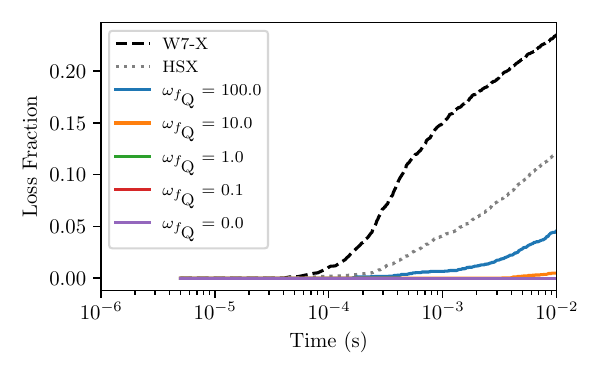}
    \includegraphics[width=.46\textwidth,
    trim={0.35cm 0.40cm 0.35cm 0.35cm}
    , clip]{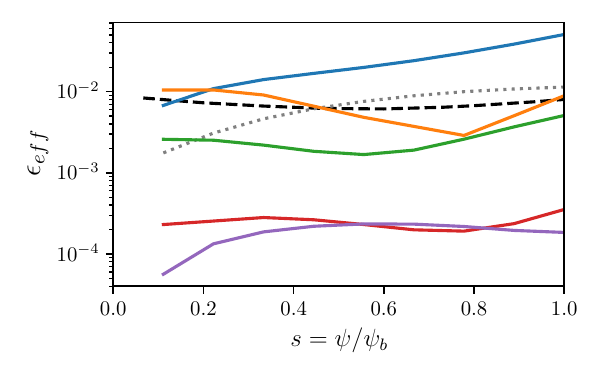}
    \caption{
    {Top: Fraction of 3.5 MeV alpha particles from a total of 3500 that are lost at $0.01$s the plasma boundary $s=1$ when started isotropically at $s=0.25$.}
    {Bottom:} Relative levels of neoclassical transport for a thermal plasma in the $1/\nu$ regime, measured using the effective helical ripple quantity $\epsilon_{\mathrm{eff}}$, between the configurations found in this work and two of the most recently built stellarators, the quasi-helical symmetric HSX machine and the quasi-isodynamic W7-X machine.
    }
    \label{fig:gx_nonlinear}
\end{figure}

As a metric for neoclassical transport, we use the 
{
parameter $\epsilon_{\text{eff}}$,
as $\epsilon_{\text{eff}}^{3/2}$
scales with the diffusive transport of trapped
particles in the long-mean-free-path collisionality regime.
}
This quantity is computed using the NEO code \cite{Nemov1999}.
In \cref{fig:gx_nonlinear} {(bottom)} we show the value of $\epsilon_{\text{eff}}$ for {the standard configuration of W7-X, HSX and the optimizations performed in this work.}
We see that the values of $\epsilon_{\text{eff}}$ for the {$\omega_{f_{\text{Q}}}=10$} microstability optimized case {is} within the range of current optimized experiments such as the HSX \cite{Anderson1995} and W7-X \cite{Beidler1990} devices{, with smaller (larger) $\epsilon_{\text{eff}}$ values for smaller (larger) $\omega_{f_{\text{Q}}}$, as expected}.
Regarding fast particle confinement, we assess in {\cref{fig:gx_nonlinear} (top)} the fraction of lost particles to the surface by tracing the guiding center motion of {3500} test particles initialized isotropically on the $s = 0.25$ surface followed using the SIMPLE code \cite{Albert2020} for {0.01}s.
For this case, the configurations are scaled to the same minor radius 1.7m and field strength on-axis of $B_0$=5.7T of the ARIES-CS fusion reactor study, in a similar fashion as in \cite{Bader2019,Landreman2022}.
We find that the optimization {cases with $\omega_{f_{\text{Q}}}$ up to $\omega_{f_{\text{Q}}}=1$} lead to no loss of particles, while {optimizations with $\omega_{f_{\text{Q}}}>1$ have finite loss fractions, with a maximum of 4.5\% at $\omega_{f_{\text{Q}}}=100$}, showing that reducing ITG growth rates in QH stellarators comes at the expense of poor particle confinement due to the degradation of quasisymmetry.

\vspace{.4cm}{\section{Conclusions}}

In this work, for the first time, microstability is taken into account in the optimization of quasisymmetric magnetic field configurations using first-principles gyrokinetic simulations.
We were able to significantly reduce the quasi-linear estimate for the heat flux, although at the cost of larger neoclassical transport.
Notwithstanding, the resulting fast particle loss is still smaller than most stellarator designs to date and with similar levels of neoclassical transport.
An extension of the framework introduced here to directly optimize for turbulent transport by replacing a quasi-linear estimate for the heat flux with its nonlinear counterpart will be the subject of future work.
{Furthermore, additional optimization targets, such as MHD stability, can be added to the objective function used, which might impose additional constraints on the magnetic field geometry.}

\vspace{.4cm}{\section{Data Availability}}

The data supporting this study's findings are available in Zenodo at https://doi.org/10.5281/zenodo.7415457.% reference number 7415457.

\vspace{.4cm}{\section{Acknowledgements}}

We thank A. Goodman for fruitful discussions as well as the GS2 and SIMSOPT teams for their invaluable contributions.
This work has been carried out within the framework of the EUROfusion Consortium, funded by the European Union via the Euratom Research and Training Programme (Grant Agreement No 101052200 — EUROfusion).  
IST activities also received financial support from FCT - Fundação para a Ciência e Tecnologia through projects UIDB/50010/2020 and UIDP/50010/2020.
R. J. is supported by the Portuguese FCT-Fundação para a Ciência e Tecnologia, under Grant 2021.02213.CEECIND and DOI  \href{https://doi.org/10.54499/2021.02213.CEECIND/CP1651/CT0004}{10.54499/2021.02213.CEECIND/CP1651/CT0004} and by the EUROfusion Enabling Research project (ENR-MOD.01.IST). 
N.R.M. was supported by the US DOE Fusion Energy Sciences Postdoctoral Research Program
administered by the Oak Ridge Institute for Science and Education (ORISE) for the DOE via Oak Ridge
Associated Universities (ORAU) under DOE contract number DE-SC0014664.
T.Q. is supported by the NSF Graduate Research Fellowship Program grant number DGE-2039656.
The optimization studies were carried out using the
EUROfusion Marconi supercomputer facility.
Views and opinions expressed here are those of the author(s) only and do not necessarily reflect those of the European Union, the European Commission, DOE, NSF, ORAU, or ORISE. None of these parties can be held responsible for them.

% \section*{References}
% \bibliographystyle{unsrt}
% \bibliographystyle{apsrev4-2}
\bibliographystyle{myunsrt}
\bibliography{references}

\end{document}